# SELF-IMMOLATIVE CHEMISTRY IN NANOMEDICINE

M. Gisbert-Garzarán[ab], M. Manzano*[ab] and M. Vallet-Regí*[ab]

[a]*Departamento de Química Inorgánica y Bioinorgánica, Facultad de Farmacia. Universidad Complutense de Madrid, Instituto de Investigación Sanitaria Hospital 12 de Octubre i + 12, Plaza de Ramón y Cajal s/n, E-28040 Madrid, Spain. E-mail: vallet@ucm.es mmanzano@ucm.es*

[b]*Networking Research Center on Bioengineering, Biomaterials and Nanomedicine (CIBER-BBN), Madrid, Spain. Fax: +34 913941786; Tel: +34 913941861*

**Abstract**

Self-Immolative Chemistry is based on the cascade of disassembling reactions triggered by the adequate stimulation and leading to the sequential release of the smaller constituent elements. This review will focus on the possibilities that this type of chemistry offers to nanomedicine research, which is an area where the stimuli responsive behavior is always targeted. There are some examples on the use of self-immolative chemistry for prodrugs or nanoparticles for drug delivery, but there is still an exciting land of opportunities waiting to be explored. This review aims at revising what has been done so far, but, most importantly, it aims at inspiring new research of self-immolative chemistry on nanomedicine.

## 1. Introduction

Nanomedicines, which have been defined as those therapeutic or imaging agents in the nanometer scale that control the biodistribution and enhance the efficacy of a therapeutic agent, have experienced a growing interest lately. Undoubtedly, cancer has been the disease in which nanomedicine has focused most of its efforts because of its unique attractive features for drug delivery, diagnosis and imaging [1]. However, the translation from the laboratory to the clinic remains challenging and only a low number of nanomedicines have been approved so far by the FDA for their clinical use [2,3].

One of the main drawbacks of current treatments is the lack of selectivity of the drugs, since those molecules are not usually capable of accumulating selectively only in the damaged areas. As a consequence, it is usual the induction of side effects in patients, such as cancer metastasis. In order to avoid that, the effort of nanotechnologists has been headed towards the development of alternative treatments in which, ideally, drugs would only affect concrete areas, thus lessening drug doses and side effects alike. Those systems are commonly known as drug delivery systems.

Most of the research carried out on nanomedicines for cancer treatment are based on therapeutic nanoparticle (NP) platforms. In this sense, NPs for drug delivery provide many advantages compared to the traditional forms of drugs, such as: control over the pharmacokinetic profile reducing the potential toxicity of a drug, protection of the therapeutic agent against potential biodegradation, the possibility of designing combined therapies loading various drugs within the same platform, and the possibility of developing targeted nanoparticles to specific tissues [4,5] or cells [6], which might be relevant when delivering cytotoxic drugs or to penetrate certain biological barriers.

Although targeted drug delivery nanosystems offer the possibility to enhance the bioavailability of drugs at the disease site, additional and interesting features can be included into nanomedicines in order to incorporate on-demand responsive behavior. In this sense, stimuli-responsive nanocarriers are able to control drug release by reacting to naturally occurring stimuli characteristic of the treated pathology, or external stimuli that can be remotely applied by the clinician (Fig. 1). The scientific literature on responsive nanocarriers is full of examples of both external stimuli, such as temperature [7], magnetic field [8], light [9] and ultrasound [10]; and internal stimuli, such as pH [11], enzymes [12] or redox species [13].

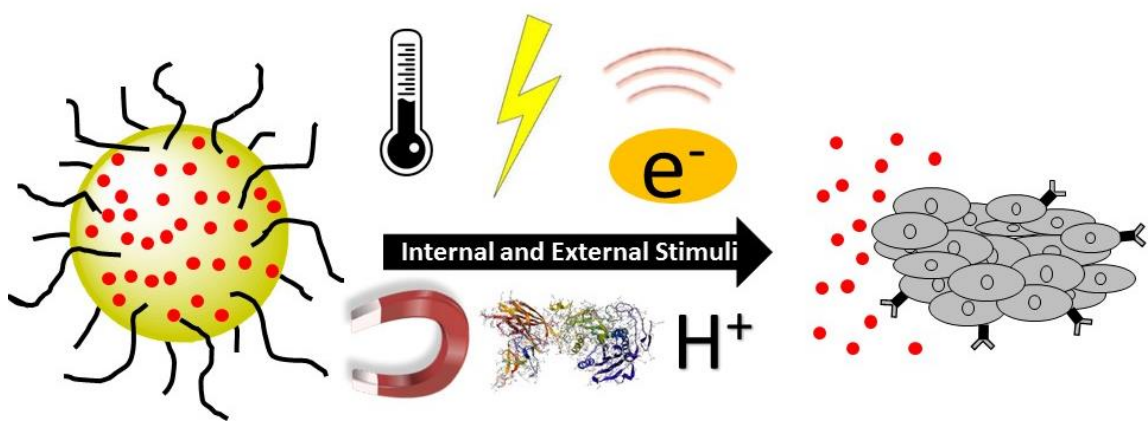

**Fig. 1.** Schematic representation of stimuli-responsive nanoparticles for drug delivery. Nanoparticle (yellow); Drug (red); Gatekeeper (e.g. polymer) (black); Tumor cells (grey).

The challenge when designing stimuli-responsive nanomedicines relies on the moiety employed to confer that responsiveness. An interesting approach consist of the so-called self-immolative systems, which are macromolecules programmed trough their synthesis to undergo head-to-tail disassembly into the building blocks in response to certain events in a domino-like fashion (Fig. 2).

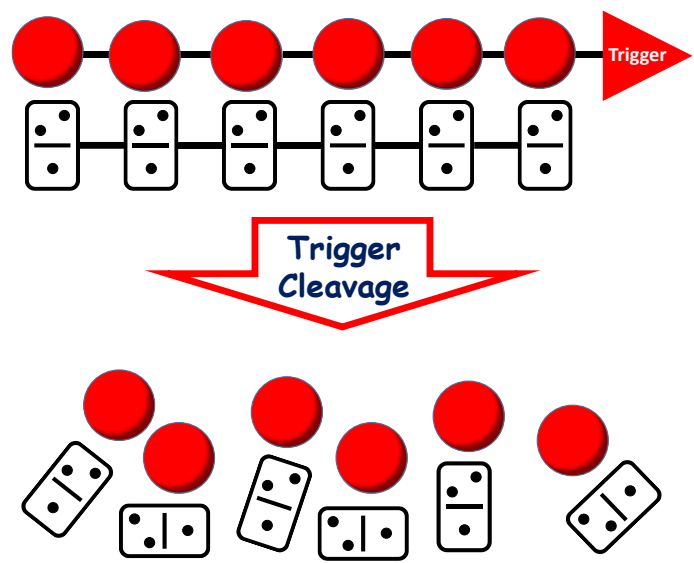

**Fig. 2.** Schematic representation of the cleavage of a trigger on a self-immolative polymer leading to the head to tail disassembly into its building blocks.

In this sense, Shabat's group laid the foundations of this new chemistry in 2003, when they reported for the first time smart self-immolative dendrimers which showed self-degradation upon stimulation [14]. Later on, they reported an interesting design based on a linear polyurethane with a protecting group on the terminal amine that acted as the trigger. Once the protecting group was removed, sequential 1,6-elimination and decarboxylation reactions led to the original monomer precursors [15].

In general, Self-Immolative Polymers (SIPs) are produced through a polymerization reaction of the appropriate monomer to then cap the terminal head-group of the polymer with a specific protecting group, leading to a polymer with a trigger. Then, the selective cleavage of the protecting group would initiate the polymer sequential fragmentation into the original building blocks from head to tail. Therefore, the appropriate selection of the trigger would allow the design of many different responsive materials.

This review will first introduce some general concepts of self-immolative chemistry, including the main properties of self-immolative spacers. Then, the application of self-immolative chemistry to nanomedicine will be revised, including some examples of prodrugs, particle-like polymeric vehicles and mesoporous silica nanoparticles capped with self-degradable gatekeepers.

## 2. **Self-Immolative Spacers**

The simplest self-immolative moiety is composed of a protecting group linked to a spacer, which is also linked to a reporter, which is finally released (Fig. 3).

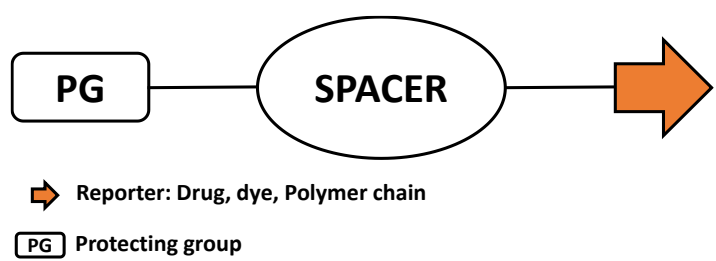

**Fig. 3.** General structure of a self-immolative spacer: A protecting group connected to a spacer which is linked to a reporter

The use of self-immolative spacers allows the release of a single molecule upon application of a certain stimulus. That stimulus breaks the bond between the protecting group (or equivalently, the trigger). After that, an electronic cascade starts leading to an intermediate species. Finally, the bond between the spacer and cargo cleaves and the reporter is released (Fig. 4).

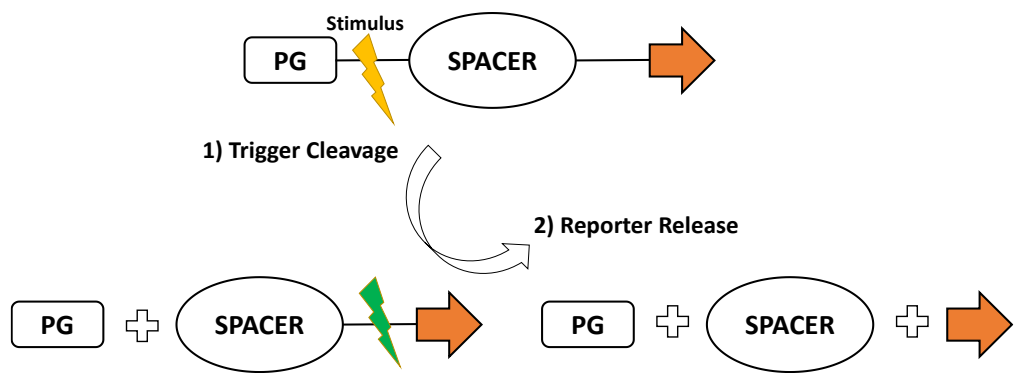

**Fig. 4.** Disassembling procedure: The protecting group (PG) is cleaved upon stimulus application (1), which finally lead to reporter release (2).

From the chemical point of view, the self-immolation process is normally driven by a cooperative rise in the entropy along with the formation of thermodynamically stable products. Generally speaking, self-immolative spacers can be classified in two main groups, depending on the particular disassembling process: (1) 1,4- ,1,6-, or 1,8-Elimination; or (2) Cyclisation [16]. Additionally, it is possible to combine both types of spacers to tune the degradation rates in self-immolative polymers [17].

a) <u>1,4-,1,6-, or 1,8-Elimination</u>

These spacers are based on aromatic species that present electron donor groups in orto or meta positions and good leaving groups in benzylic positions. The electron donor group, which is usually chosen among amino, hydroxyl or thiol functionalities, normally leads to 1,4- or 1,6-Eliminations. Thus, once the self-immolation process starts trough the cleavage of the protecting group, the electron donor group transmit the electron pair to the aromatic ring, starting an electronic cascade that leads to the disassembly of the spacer (Fig. 5).

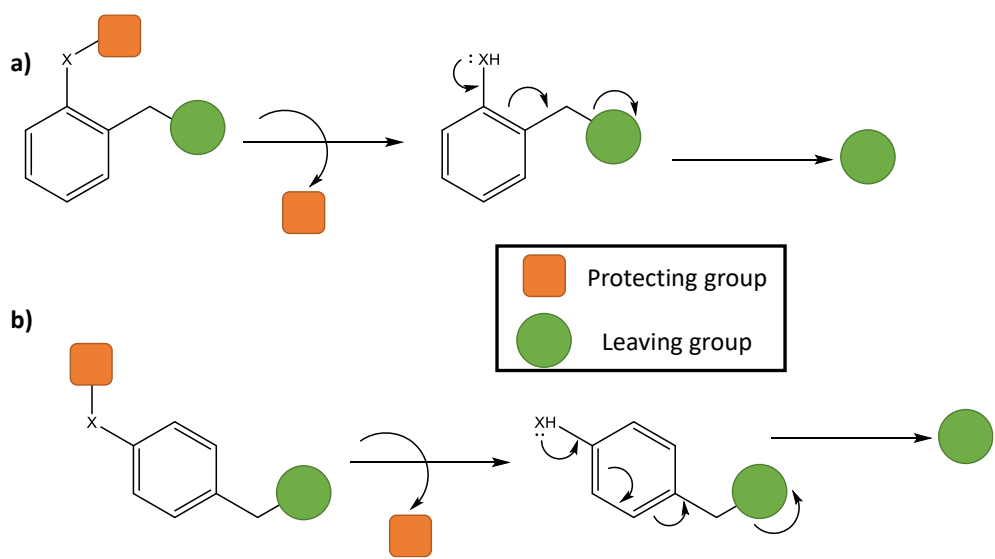

**Fig. 5.** Spacers and elimination procedures. a) 1,4-Elimination; b) 1,6-Elimination. When the protecting group is cleaved, an active nucleophile is generated, which donate the electron pair and initiates the self-immolation.

Interestingly, 1,8-Elimination does not lead to the release of the leaving group in molecules such as naphthalene. Moreover, 1,10-Elimination or successive do not provide the cargo. This behavior has not been completely elucidated although some explanations have been given. It has been associated to the excessive energy barrier that is necessary to break aromaticity and to the repulsion of the ortho-hydrogens that prevents the formation of the planar structure necessary for good electron delocalization [16].

b) Cyclisation

Spacers suffering cyclisation are based on lineal alkyl chains or aromatic molecules with substituents in ortho. Once the trigger would be cleaved, a nucleophilic attack on a carbonyl group or an electrophilic aliphatic carbon leads to the cyclisation of the spacer, which provokes the release of the reporter [16]. Both examples are shown in Fig. 6.

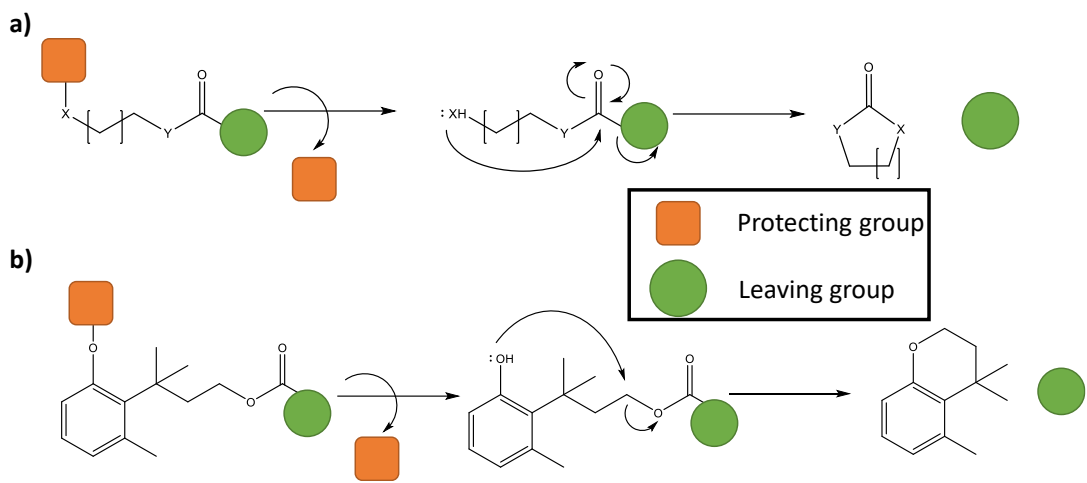

**Fig. 6.** Spacers suffering cyclisation. a) Attack to carbonyl; b) Attack to aliphatic carbon.

A complete and exhaustive description of self-immolative spacers can be found elsewhere [16,18–21].

The election of the protecting group emerges as a powerful tool to control the on-demand beginning of the self-immolation. A number of protecting groups useful in a clinical scenario have been reported. Those stimuli can be classified in three main groups: (1) chemical (hydrogen peroxide or pH); (2) enzymatic; and (3) luminescent (ultraviolet or near-infrared light). Table 1 summarizes the different types of stimuli-responsive protecting groups.

**Table 1.** Stimuli-responsive protecting groups for self-immolative spacers

| Protecting Group | Trigger | References |
|---|---|---|
| **Chemical Stimulus** | | |
| [pinacol boronate benzyl ether structure] | $H_2O_2$ | [22–27] |
| [Boc carbamate structure] | Acid pH | [28–31] |
| 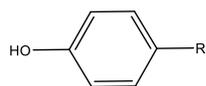 | Basic pH | [32] |
| 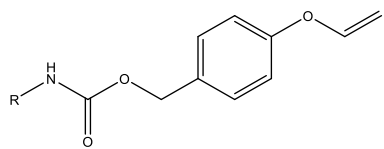 | Tetrazine | [33] |
| **Enzymatic Stimulus** | | |
| 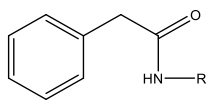 | Penicillin G-Amidase | [34] |
| 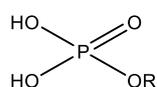 | Alkaline Phosphatase | [35] |
| 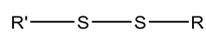 | Glutathione | [22,36–41] |
| 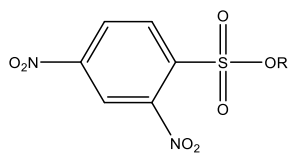 | Glutathione | [42] |
| 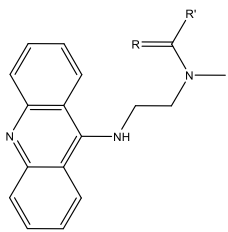 | Human carboylesterase-2 | [43] |
| 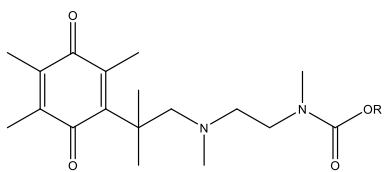 | DT-Diaphorase | [44] |
| **Light Stimulus** | | |
| 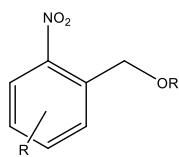 | Ultraviolet | [24,41,45–51] |

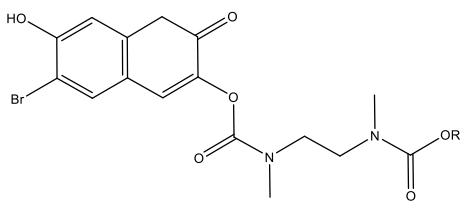

| | Near-Infrared | [50–52] |

---

## 3. Self-Immolative chemistry in nanomedicine

### 3.1. Prodrugs based on self-immolative spacers

The term *prodrug* refers to masked forms of active drugs. These compounds are supposed to be inactive or to have negligible therapeutic activity. In order to recover it, prodrugs must be converted into the active drug. That process usually take place either through a chemical or enzymatic process. These structures are usually designed to increase the selectivity of the treatment so that doses can be reduced [53]. It is also usual the design of prodrug-like precursors that are capable of releasing dye molecules. By doing that, it is possible to monitor different processes. In this sense, the detection of a light signal indicates that the molecule has self-immolated.

For an enzymatic reaction to happen it is necessary the formation of temporary bonds between the substrate and the active site of the enzyme. Then, if an enzyme is to bind a substrate, the later should show adequate accessibility for the active site. In that sense, it has been shown that the addition of a self-immolative spacer between the trigger and the deliverable drug increases the accessibility of the enzyme to the specific substrate. Then, the introduction of self-immolative fragments leads to higher rates of bioconversion and, then, higher therapeutic activity [54]. Therefore, the prodrug will travel throughout the body without therapeutical activity until arriving at the desired area, where it will suffer self-immolation and it will recover its therapeutical activity.

There are several examples of enzymes that have been utilized in the design of this kind of molecules. Cathepsin B is an cysteine protease which is involved in many pathologies and oncogenic processes in humans [55]. For instance, it is overexpressed in the lysosomes of many cancer cells. The self-immolative spacer p-aminobenzyl alcohol has been widely used in combination with various combinations of dipeptides to provide sensitivity to this enzyme. For instance, the dipeptides Phe-Lys and Lys-Lys have been used in the design of selective prodrug-inspired probes capable of releasing a fluorogenic peptide [56]. Besides, the Phe-Lys combination has shown great efficacy both *in vitro* and *in vivo*, showing much less toxicity in healthy areas than its active counterpart [57]. In order to monitor endogenous alkaline phosphatase in living cells, p-hydroxybenzyl alcohol has been used as spacer between a phosphate moiety (enzyme substrate) and the dye resorufin [35]. DT-diaphorase, an enzyme that is overexpressed in some malignant tumor, has also been used as enzymatic trigger through self-immolative linkers. In particular, the use of a quinone propionic acid-based moiety allows to incorporate two camptothecin drug molecules at a time on each prodrug [44]. Furthermore, there is an enzyme (human carboxyl-esterase-2) that has proved its efficacy in the cleavage of a self-immolative ester to provide the active form of the potent anticancer drug platinum-acridine [43].

With regard to self-immolative prodrugs undergoing chemical activation, the majority of the examples reported involves a conversion mechanism based on the action of the tripeptide glutathione (GSH), although designs sensitive to hydrogen peroxide have also been reported [22]. GSH, which is overexpressed in various types of tumors [58,59], is capable of reducing disulfide bonds, as those present in disulfide-based self-immolative linkers [60]. Besides, conjugation of self-immolative prodrugs with gold nanoparticles has also been described, leading to a system with dual drug capability and targeting ability [37]. In order to verify the effective drug cleavage due to the action of GSH, a dendritic system bearing one molecule of camptothecin and one of inactive dye was reported. Then, when GSH cleaves the trigger, both drug and dye are released in their active form, thereby obtaining a light signal only if the drug has indeed been released [42].

So far, the examples shown usually involve two drug molecules maximum, however, it might be interesting the use of polymer chains bearing pendant cleavable disulfide-based prodrugs, so more drug molecules could be delivered. In particular, this approximation has been found to be successful in treating HIV, since it is possible to carry multiple drugs at a time thus leading to a more effective combination therapy [36,61]. Following a similar approximation, a polyethylene glycol block bearing a small amount of prodrugs (doxorubicin) on its side chain has been reported. In this case, the addition of tetrazine leads to almost complete conversion into the free drug thanks to the conversion of a phenyl vinyl ether into a phenol [33].

### 3.2. Nanocarriers based on self-immolative moieties

As it has been shown above, it is very frequent the use of single self-immolative units in order to design prodrug-like molecules. Besides, it is also possible to design several types of nanovehicles by using self-immolative fragments in combination with other macromolecular structures. In particular, we will provide here an overview on polymer capsules, vesicles, micelles and nucleic acid-nanoparticles as well. Moreover, we will take a look to the unexplored domain of using self-immolative chemistry to hamper undesired drug release from mesoporous silica nanoparticles.

**Polymer capsules**

In the last few years a number of techniques to synthesize particles or capsules out of self-immolative elements have been developed [62]. In that sense, the synthesis of self-immolative microcapsules composed of self-immolative polymers was accomplished in 2010 for the first time [31]. The presence of the triggering units throughout the wall led to the disruption of the carrier and the release of the cargo upon triggering removal conditions. Afterwards, other design of self-immolative polymer was formulated as nanoparticles. In this case, the nanocarrier was sensitive to both ultraviolet and near-infrared light (instead of to strong acid or base, as in the first example) and then, the applicability of these nanocarriers to biological scenarios was finally accomplished [50]. Furthermore, more examples comprising other designs of self-immolative polymers sensitive to various wavelengths have also been reported [49,51,52]. The introduction of aryl boronic esters within the polymer chain has led to multiple designs of hydrogen peroxide responsive polymer nanoparticles [26,27,63]. Moreover, dual stimuli-responsive polymer nanoparticles have also been published, including pH and light [29,64] or hydrogen peroxide and pH [25]. Finally, heat-responsive polymer nanoparticles have been recently reported. Upon heating, the carrier undergoes a retro-Diels Alder reaction that leads to a furfuril carbamate that suffers elimination, thus provoking the degradation of the carrier [65].

**Polymeric micelles and vesicles**

It is also possible to design vehicles that self-assemble spontaneously from linear structures to three dimensional polymeric carriers.

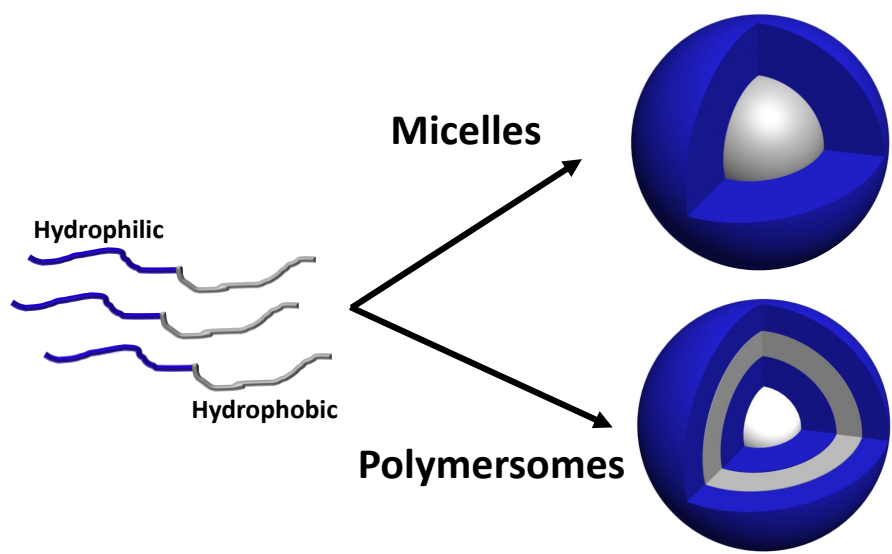

**Fig. 7.** Cross-section of the Self-Assembly of amphiphilic block copolymers. Top: Formation of micelles; Bottom: Formation of polymersomes.

Polymeric micelles (Fig. 7, top) are core-shell structures formed by amphiphilic block copolymers in aqueous solution, with the hydrophobic block forming the core and the hydrophilic the shell. Depending on the polymer composition and the preparation, those copolymers can form vesicular assemblies, the so-called polymersomes. Polymersomes (Fig. 7, bottom) are similar to liposomes, since they show a bilayer structure with an aqueous inner cavity capable of hosting hydrophilic drugs [66]. In particular, for a fixed hydrophilic part, the increase of the size of the hydrophobic section leads first to micelles and then, worms, vesicles and tubular vesicles respectively [67].

The disassembly of this kind of nanocarriers usually involves changes in the solubility of the hydrophobic block, which demand a cooperative behavior of all monomer units. For that reason, it is particularly interesting the use of self-immolative components in order to obtain that cooperative behavior during disassembly. These self-destructive fragments allow the complete disintegration of the vehicles only in the desired area, thereby releasing selectively the cargo.

Various approximations have been proposed in order to construct self-immolative polymeric micelles. The general design normally comprises two polyethylene glycol blocks having a self-immolative construction within them. Polyethylene glycol is interesting since not only provides a hydrophilic segment to the block copolymer, but also provides the carrier with improved stealth properties. It is usual the use of a self-immolative block containing multiple triggering points along the chain. In that sense, hydrogen peroxide-responsive poly(amino ester)s [48] as well as disulfide containing polycarbonates [47] have been used as building blocks for the hydrophobic component. Nonetheless, in order to reduce the triggering events, and therefore, increase the efficacy of the self-immolative process, it would be desirable to have only one triggering unit. To achieve that, micelles formed from amphiphilic polymers containing a poly(ethyl glyoxylate) undergoing head-to-tail depolymerization have been also synthesized [24].

The inclusion of pendant self-immolative groups [46] or linear self-immolative polycarbamates [41] within hydrophilic segments has led to light-responsive polymersomes. In that sense, self-immolative polymersomes can be decorated with targeting peptides to improve the therapeutic efficacy of the carrier [23]. Moreover, enzyme-responsive self-immolative polymeric vesicles capable of addressing antimicrobial resistance have shown their efficacy *in vitro* [34].

**Nucleic acid-based nanoparticles**

It has been observed that there are particular three-dimensional arrangements of nucleic acids with exquisite properties that their linear or circular counterparts do not show. For instance, those structures show better binding with complementary strands, stability against enzyme degradation and internalization. Although it is common the coating of solid gold nanoparticles with DNA strands, it is also possible to design vehicles entirely made of nucleic acids. In that sense, DNA micelles bearing light [45] and reductive [40] self-immolative moieties have been reported. Moreover, it is also possible the combination of miRNA with polymeric prodrugs [39] or polymer chains bearing pendant prodrugs [38] to give rise to particle-like polyplexes. The main advantage of all these designs is that it is possible to transport drugs, in their active or inactive form, that, once the stimulus is applied, are released in combination with different nucleic acids, which also show therapeutic efficacy and biocompatibility.

### 3.3. Self-immolative moieties as gatekeepers for mesoporous silica nanoparticles

Among all the nanoparticles that are being investigated as potential nanomedicines, mesoporous silica nanoparticles (MSNs) have drawn the attention of the scientific community because of their attractive properties to be used as drug delivery nanosystems [68–72]. Basically, they are very robust, because of their silica network constitution; they provide a high loading capacity, thanks to the network of cavities that can be loaded with many different therapeutic agents; and they are very easy to chemically modify, thanks to the silanol groups present on their surface.

The synthesis of MSNs is based on the sol-gel process, through the hydrolysis and condensation of the silica precursors, which will then form a network of silica

[73]. The mesoporous structure of those nanoparticles is achieved thanks to the use of surfactants as structure directing agents during the synthetic process. Then, once the process is finished, removing the surfactant would lead to the network of cavities typical from these materials. Finally, to obtain nanoparticle, the synthetic process needs to be carried out under very dilute conditions, following a modification of the sol-gel process. Following this protocol, spherical nanoparticles with diameters between 100 and 200 nm and empty pores of *ca.* 2 nm can be obtained (Fig. 8).

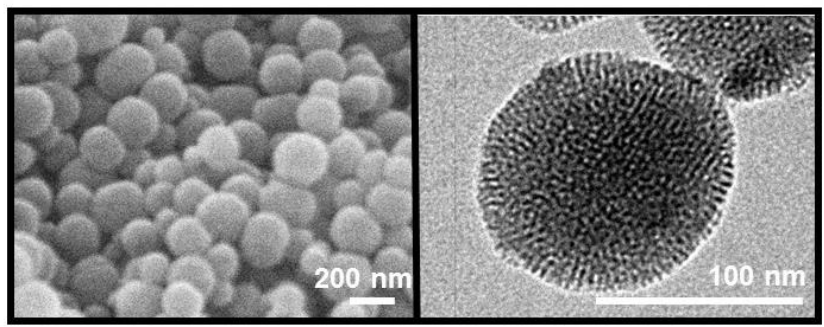

**Fig. 8.** Scanning Electron Microscopy (left) and Transmission Electron Microscopy (right) images of Mesoporous Silica nanoparticles.

When using MSNs as drug delivery vectors, there is an important aspect that needs to be considered: they are materials with open porosity, which means that it is very easy to introduce drugs into their pores, but it is also very easy for those drugs to diffuse out. Consequently, the pore entrances should be closed to avoid the premature release of the cargo, but those gates need to be designed so they can be opened on command to release the cargo. This is precisely the way that stimuli-responsive nanocarriers work: they release their content on-command in response to certain stimuli, which can be internal and characteristic of the treated pathology or external and remotely applied by the clinician.

The scientific literature is full of examples of MSNs able to respond to different stimuli, and among them, systems take advantage of the significant variations in pH (gastrointestinal track, tumor microenvironment or endosomal/lysosomal compartments of the cell) are very attractive [11]. In this sense, cancer cells normally use aerobic glycolysis as their source of energy, which leads to lactate accumulation within the tumor microenvironment that lowers the pH. Thus, employing functional moieties able to respond to pH variations as gatekeepers on MSNs seems an interesting approach to specifically release therapeutic agents on those sites. Up to date, the combination of MSNs with self-immolative molecules sensitive to pH variations has been explored from 2 perspectives: in response to acid pH [28] or to basic pH [32].

Acid pH responsive nanocarriers have been prepared through the combination of MSNs and self-immolative polymers in a unique carrier that is able to combine the benefits of both components: the robustness and great loading capacity of MSNs and the responsive behavior of SIPs. At physiological pH the SIP would maintain its structure while at acid pH it would start to disassemble, thus opening the pores and allowing the release of the cargo (Fig. 9).

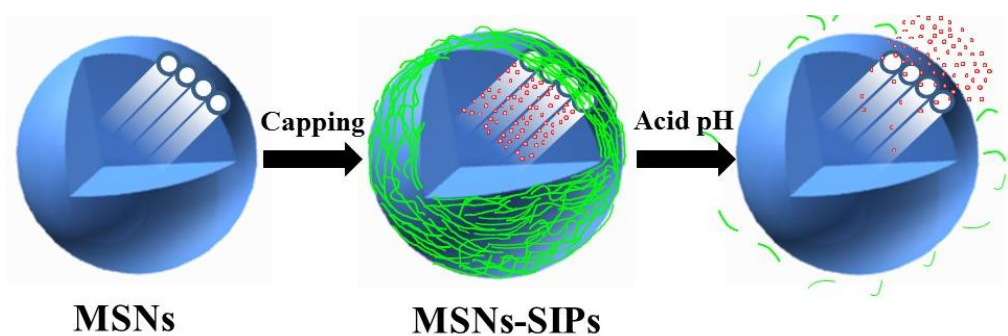

**Fig. 9.** Schematic representation of a pH sensitive nanocarrier composed of Mesoporous Silica Nanoparticles and Self-Immolative Polymers.

The self-immolative polymer was based on a polyurethane backbone with a tert-butyloxycarbonyl protecting group acting as the trigger. The polymer was produced using a previously synthesized monomer from 4-aminobenzylalcohol and phenyl chloroformate, and polymerized using a tin catalyst. The trigger sensitive to acid pH was based on a tert-butyl carbamate. Once the polymer was characterized and the self-immolative behavior was observed, it was grafted to the surface of MSNs in two stages. First, the nanoparticles were funcionalised with a linker through silanol chemistry to then anchor the polymer to the other end of the linker. The successful grafting of SIP on MSNs was characterized through several techniques and the cargo release under different pHs was evaluated using a fluorescent dye as a model molecule. (Fig. 10)

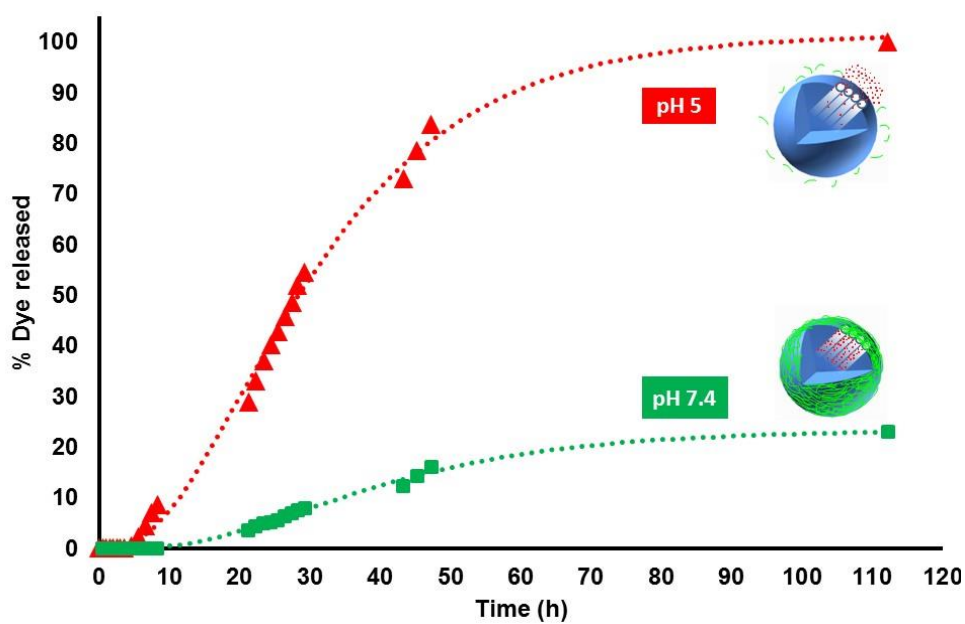

**Fig. 10.** Fluorescent dye cumulative release at physiological pH and acid pH.

At physiological pH there was a very small amount of dye released, probably due to the dye molecules entrapped into the polymer network. However, when the pH of the solution was acid, a great release of the cargo was observed, confirming the disassembling of the polymer and allowing the dye molecules to diffuse out of the pores.

On the other hand, the responsive system to basic pH was produced through the capping of MSNs with self-immolative molecular gates [32]. That molecular gate was a carbamate derivative that could undergo a 1,6-elimination reaction through a quinone-methide cascade under basic conditions. The self-immolative molecule was anchored to the surface of MSNs through silanol chemistry after the cargo molecules (sulforhodamine B) were introduced into the network of cavities of the nanocarriers. The cargo release from the designed carriers was evaluated at different pHs, observing that at acid and neutral pHs it was highly inhibited. However, at pH 8 the release reached almost 100% after 120 minutes, which was indicative that the self-immolative process took place, inducing the rupture of the gate and promoting the cargo release.

### 3.4 Biocompatibility and cytotoxicity

It has been shown that they by-products of the self-immolation could interact with enzymes (that trigger the self-immolation) [74]. However, there is still a lack of research regarding the study of the biocompatibility and cytotoxicity of the by-products of the self-immolation. In general, the studies shown here do not study the effect of those residues. These nanocarriers have been shown to be biocompatible when the stimulus is not applied, thus showing cytotoxicity only upon stimulation. Since the stimuli are only applied in the tumor cells, the side effects that might be caused by the by-products would only have sort of a synergistic effect with the cytotoxic drug. In conclusion, more research on the possible negative effects should be carried out.

### 4. Conclusions

The most important concepts of Self-Immolative Chemistry have been here presented, highlighting the importance of the trigger to initiate the disassembling process and the spacer for allowing the cascade reactions. This type of materials developed by Shabat´s group offer very attractive features to be used on nanomedicine research, especially in stimuli-responsive systems. This review has presented some representative examples but we honestly consider that there are many different possibilities and unexplored opportunities for this type of Self-Immolative Chemistry in the area of Nanomedicine.

**Acknowledgments:** The authors thank funding from the EU H2020-NMP-PILOTS-2015 program through Grant No. 685872 (MOZART) and the European Research Council (Advanced Grant VERDI; ERC-2015-AdG Proposal No. 694160). For the elaboration of the graphical abstract a template from http://www.servier.com/Powerpoint-image-bank has been used.